\documentclass[aps,prl,twocolumn,amsmath,amssymb]{revtex4}
\usepackage{graphicx}

\newcommand{\hp}{(C$_5$H$_{12}$N)$_2$CuBr$_4$}
\newcommand{\CuPzN}{copper pyrazine dinitrate}
\newcommand{\kp}{\kappa^\|}
\newcommand{\kq}{\kappa^\perp}

\begin{document}

\bibliographystyle{prsty}

\title {Spinon localization in the heat transport of the spin-1/2 ladder compound (C$_5$H$_{12}$N)$_2$CuBr$_4$}

\author{ A. V. Sologubenko,$^1$   T. Lorenz,$^1$ J. A.
Mydosh,$^1$\footnote{Present address: Max Planck Institute for
Chemical Physics of Solids, 01187 Dresden, Germany}
    B. Thielemann,$^2$ H. M. R{\o}nnow,$^3$
    Ch. R\"{u}egg,$^4$ and K. Kr\"{a}mer$^5$}

\affiliation{$^1$II. Physikalisches Institut, Universit\"{a}t zu
K\"{o}ln, Z{\"u}lpicher Str.\ 77, 50937 K\"{o}ln,  Germany \\
$^2$Laboratory for neutron scattering, ETH Z\"{u}rich \& Paul
Scherrer Institute, 5232 Villingen PSI, Switzerland \\
$^3$ Laboratory for Quantum Magnetism, \'{E}cole Polytechnique
F\'{e}d\'{e}rale de Lausanne (EPFL), Switzerland \\
$^4$London Centre for Nanotechnology and Department of Physics and
Astronomy, University College London, London WC1E 6BT, United
Kingdom \\
$^5$Department of Chemistry and Biochemistry, University
of Bern, Freiestrasse, CHÐ3000 Bern 9, Switzerland}

\email{lorenz@ph2.uni-koeln.de}

\date{\today}

\begin{abstract}
We present experiments on the magnetic field-dependent thermal
transport in the spin-1/2 ladder system
(C$_5$H$_{12}$N)$_2$CuBr$_4$. The thermal conductivity $\kappa(B)$
is only weakly affected by the field-induced transitions between
the gapless Luttinger-liquid state realized for $B_{c1}< B <
B_{c2}$ and the gapped states, suggesting the absence of a direct
contribution of the spin excitations to the heat transport. We
observe, however, that the thermal conductivity is strongly
suppressed by the magnetic field deeply within the
Luttinger-liquid state. These surprising observations are
discussed in terms of localization of spinons within finite ladder
segments and spinon-phonon umklapp scattering of the predominantly
phononic heat transport.\end{abstract}

\maketitle

Studies of the heat transport in 1D spin systems are of strong
current
interest.~\cite{HeidrichMeisner07,Sologubenko07_JLTP,Hess07_EPJ}
From the theoretical side there is consensus that the intrinsic
spin-mediated heat transport of integrable spin models is
ballistic, while the situation in nonintegrable spin models is
less clear. Experimentally, such studies were stimulated by the
observation of a strong anisotropy of the thermal conductivity
$\kappa(T)$ in the spin-1/2 ladder compound
(Sr,Ca,La)$_{14}$Cu$_{24}$O$_{41}$, which has been explained by a
large spin contribution $\kappa_s$ along the ladder
direction~\cite{Sologubenko00_lad,Hess01,Kudo01}. In order to
relate model calculations to experimental data, theory has to
incorporate the coupling between spin excitations and the
underlying lattice, while experimentally it is necessary to
separate $\kappa_s$ from the measured total $\kappa$. Here,
studies of the magnetic-field dependent $\kappa(B,T)$ can provide
much more information than just the zero-field $\kappa(T)$, since
strong enough magnetic fields change the spin excitation spectra
and cause transitions between different quantum phases. Yet, the
above-mentioned (Sr,Ca,La)$_{14}$Cu$_{24}$O$_{41}$ is unsuitable
for such studies due to the strong exchange interaction
($\sim$~2000~K) in the ladders that requires magnetic fields far
above typical laboratory magnets.

Piperidinium copper bromide
(C$_5$H$_{12}$N)$_2$CuBr$_4$~\cite{Patyal90} is one of the best
spin-1/2 ladder compounds with weak intraladder exchange. It has a
monoclinic structure (space group P2$_1$/c) with the ladders
running along the $a$ axis. The rung exchange $J_\perp\simeq 13$~K
is about four times larger than the leg exchange
$J_\parallel\simeq 3.6$~K
\cite{Watson01,Lorenz08,Anfuso08,Klanjsek08,Ruegg08cm,Thielemann08cm,Thielemann09PRL}.
The zero-field spin excitation gap $\Delta\simeq 9.5$~K decreases
in an external magnetic field, so that between $B_{c1} \simeq
7.1$~T and the full saturation field $B_{c2} \simeq 14.5$~T (for
$B \parallel a$ where $g=2.06$ \cite{Patyal90}) the spin
excitations are gapless and the system is in the Luttinger liquid
(LL) state. The LL state extends down to the temperature of a 3D
ordering $T_N \leq 110$~mK suggesting an interladder coupling $J'
\sim 27\,{\rm mK} \ll J_\perp,
J_\parallel$~\cite{Klanjsek08,Ruegg08cm,Thielemann08cm}.

Here, we present measurements of the thermal conductivity of \hp\
between 0.3 and 10~K in magnetic fields up to 17~T. Surprisingly,
no sign of spin-mediated heat transport is observed, which is
interpreted by spinon localization. The heat transport is
dominated by phonons being strongly scattered by spin excitations
in the vicinity of the commensurate wavevector.

Two samples of \hp\ of approximate dimensions $5 \times 1.7 \times
0.8$~mm$^3$ were cut from two crystals of the same batch with the
longest dimension either along the $a$ axis (along the ladders) or
along the $c^*$ axis. The thermal conductivity was measured using
the standard uniaxial heat-flow method where the temperature
difference was produced by a heater attached to one end of the
sample and monitored using a matched pair of RuO$_2$ thermometers.
The heat flow was directed along the longest direction of each
sample and the magnetic field was parallel to $a$.

\begin{figure}[t]
   \begin{center}
    \includegraphics[width= 8.3cm]{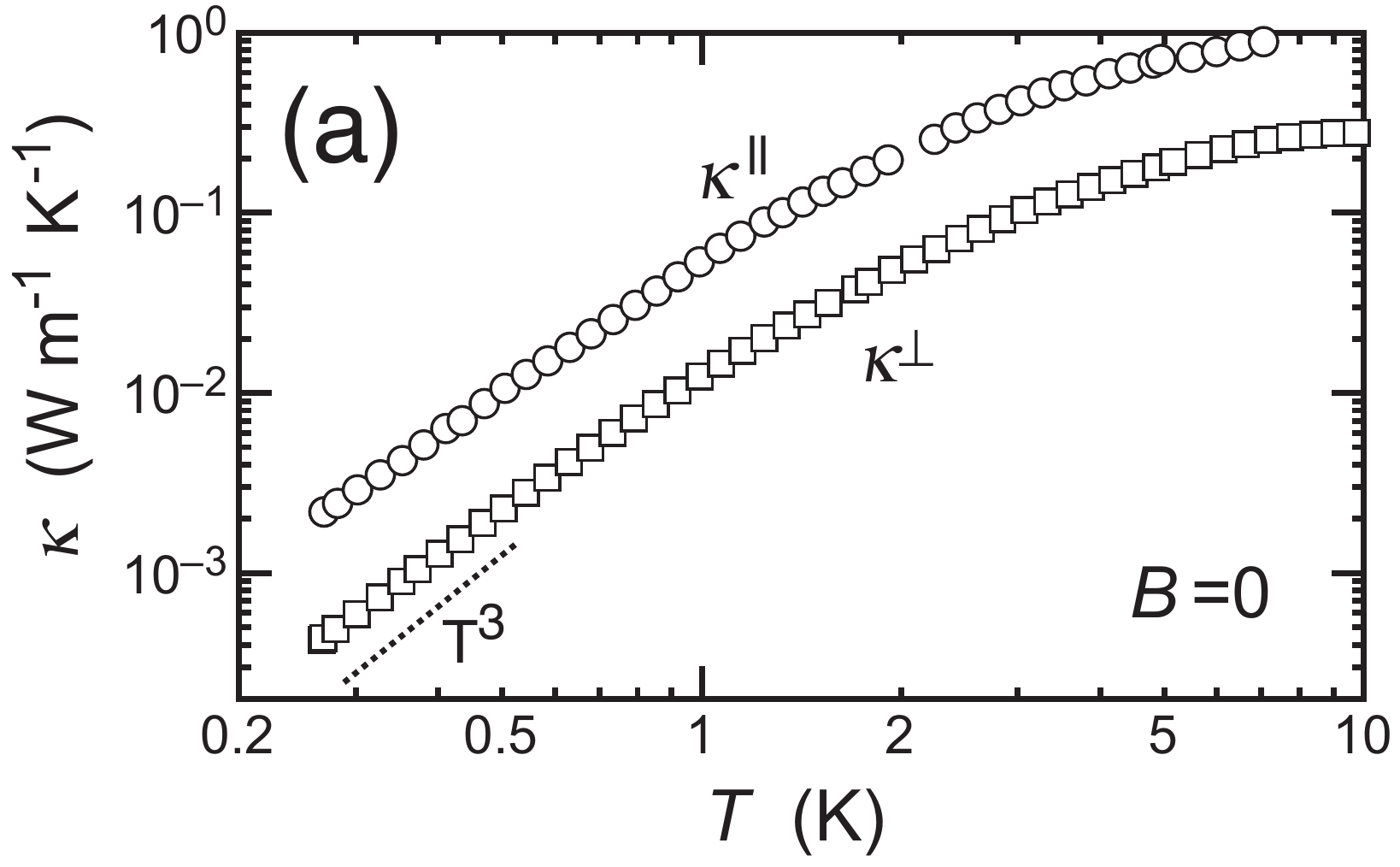}
    \includegraphics[width= 8.3cm]{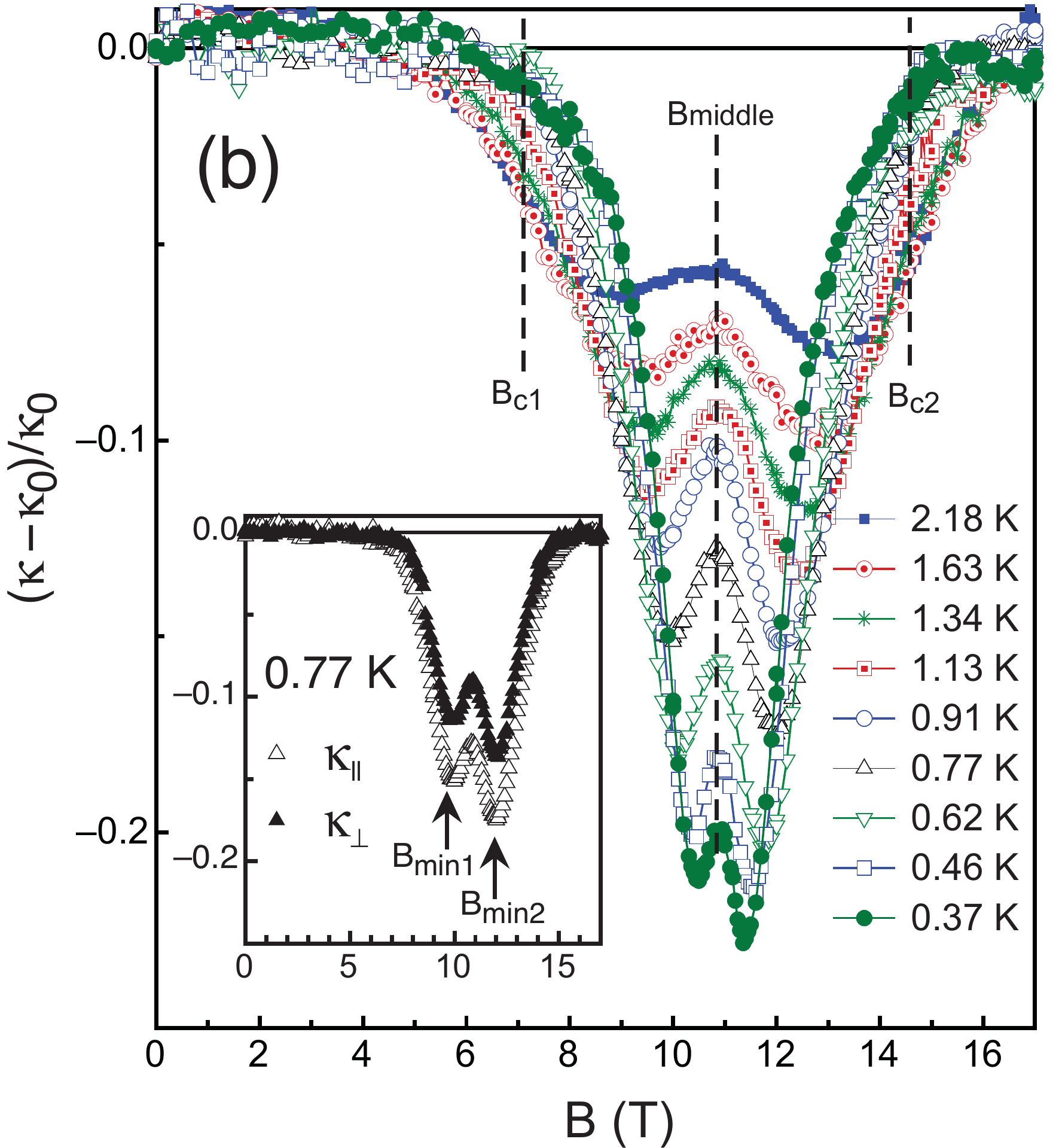}
     \caption{(color online)
     (a ) Zero-field thermal conductivity of \hp\ parallel and perpendicular
     to the ladder direction as a function of temperature.
     (b) Relative change of the thermal conductivity parallel to
     the ladders as a function of magnetic field at constant temperatures.
     Inset: Data for $T=0.77$~K parallel and perpendicular to the ladders.
    }
\label{KTH}
\end{center}
\end{figure}

The zero-field thermal conductivities $\kp,\kq$ for both
directions of the heat flow are shown in Fig.~\ref{KTH}~(a). Their
behavior is typical for a phononic heat transport, approaching
$\kappa \propto T^3$ at lowest temperature. The field dependencies
of the thermal conductivity normalized to its zero-field value at
several constant temperatures are presented in Fig.~\ref{KTH}~(b).
The most pronounced feature of the $\kappa(B)$ curves are two
minima located in the middle of the field interval between
$B_{c1}$ and $B_{c2}$ symmetrically with respect to $B_{\rm
middle}=(B_{c1}$ + $B_{c2})/2 \simeq 10.8$~T. With decreasing
temperature, the two minima become deeper and move closer to
$B_{\rm middle}$. It is noteworthy that the critical fields
$B_{c1}$ and $B_{c2}$ are not marked by a distinct feature of
$\kappa(B)$.  As displayed in the inset of Fig.~\ref{KTH}~(b),
$\kp (B)$ and $\kq (B)$ show the same behavior.

The total heat conductivity of a magnetic insulator comprises the
phonon and the spin contributions
\begin{equation}
\label{eKtotal}
\kappa = \kappa_s + \kappa_{\rm ph},
\end{equation}
where each contribution is given, as a first approximation,  by a
product of the specific heat $C_{i}$, the average velocity $v_i$,
and the mean free path $\ell_i$ of phonons ($i={\rm ph}$) and spin
excitations ($i=s$). In addition, a term in Eq.~(\ref{eKtotal})
due to spin-phonon drag is expected \cite{Boulat07}. In quasi-1D
systems, both  $\kappa_s$ and spin-phonon drag are only essential
along the chain (ladder) direction. From the almost absent
anisotropy of $\kp (B)$ and $\kq (B)$ we conclude that $\kappa_s$
and the drag contribution are small.

The spin contribution $\kappa_s(T,B)$ can be analyzed using a
mapping of the spin-1/2 ladder onto the effective spin-1/2 XXZ
chain. Within this mapping, the ground state singlet and the
low-energy component of the triplet form an effective spin
$\tilde{S}=1/2$  on each rung of a ladder
\cite{Totsuka98,Mila98,Chaboussant98}. The effective Hamiltonian
is
\begin{eqnarray}
\label{eEffHamiltonian}
    H_{XXZ} & =& J \sum_{i}^{N}  \left( \tilde{S}_x^{i} \tilde{S}_x^{i+1} + \tilde{S}_y^{i} \tilde{S}_y^{i+1} + \delta \tilde{S}_z^{i} \tilde{S}_z^{i+1}   \right) \nonumber\\
 & - & g \mu_B \tilde{B} \sum_{i}^{N} \tilde{S}_z^i,
\end{eqnarray}
where $\delta$ is the anisotropy parameter. For a spin-1/2 ladder
in the strong-coupling limit ($J_{\perp}/J_{\parallel} \gg 1$),
$\delta= 1/2$,  $J = J_\parallel$, and $\tilde{B} =  B  - (J_\perp
+ J_\parallel/2)/g \mu_B$. Thus, $B=B_{\rm middle}$ corresponds to
the zero effective field $\tilde{B} = 0$. The model
hamiltonian~(\ref{eEffHamiltonian}) well describes many observed
low-temperature features of \hp, as e.\ g.\ the magnetization
\cite{Watson01}, thermal expansion \cite{Lorenz08}, NMR
\cite{Klanjsek08}, and specific heat \cite{Ruegg08cm}
measurements. Some discrepancies are mainly caused by the fact
that the condition $J_{\perp}/J_{\parallel} \gg 1$ is only
approximately satisfied.

\begin{figure}[t]
   \begin{center}
     \includegraphics[width= 8.6cm]{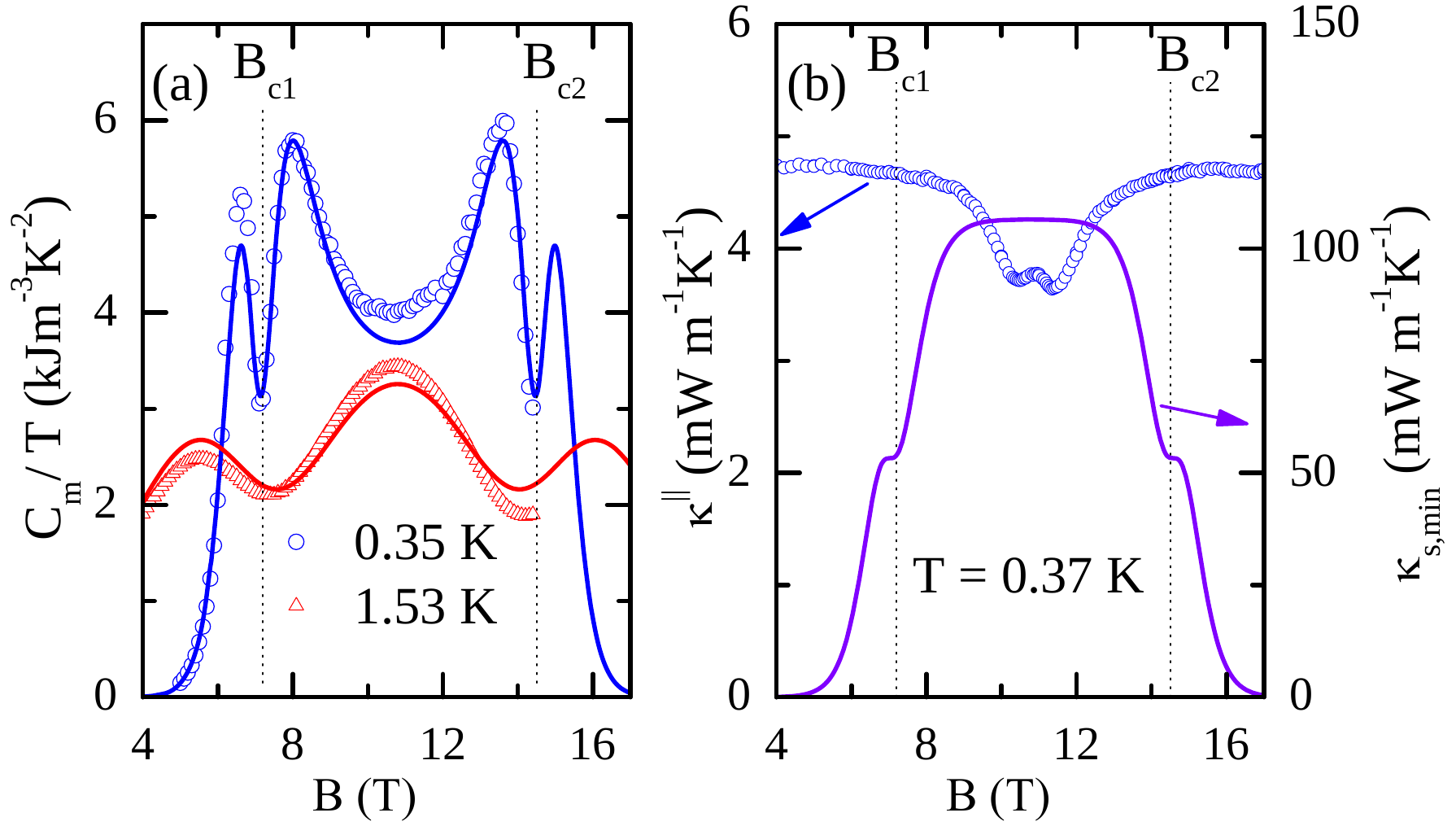}
     \caption{(color online)(a) $B$-dependence of the magnetic specific heat of \hp .
     The symbols are experimental data~\cite{Ruegg08cm} and the lines are
     calculated within the MFT model (see text).
    (b) Comparison of the expected spin thermal conductivity
     $\kappa_{s,{\rm min}}$ (line) and the experimental
     total thermal conductivity $\kappa^{\parallel}$ along the ladders
     (symbols).}
\label{CmKmMFT}
\end{center}
\end{figure}

The heat transport in the spin-1/2 XXZ chain can be well described
for $T < J/k_{\rm B} $ within a relaxation time approximation
combined with a mean-field theory (MFT) approach via the
Jordan-Wigner (JW) transformation. Eq.~(\ref{eEffHamiltonian}) is
mapped onto a system of interacting spinless fermions
\cite{HeidrichMeisner05,HeidrichMeisner05PhD}, which  occupy a
cosine band
\begin{equation}\label{eMFTdisp}
\varepsilon_k = - J (1 + 2 \delta \Omega) \cos(ka) - g \mu_B
\tilde{B} + 2 J \delta m \,.
\end{equation}
Here, $k$ is the wave vector, $m$ the magnetization, and  $a$ the
distance between neighboring spins. The parameters $\Omega$ and
$m$ are determined from  $\Omega =
\frac{a}{\pi}\int_{0}^{\pi/a}{\cos(ka) f_k dk }$ and $ m =
-\frac{1}{2} + \frac{a}{\pi} \int_{0}^{\pi/a}{f_k dk },$ where
$f_k = (\exp(\varepsilon_k/k_B T)+1)^{-1}$ is the Fermi
distribution function. The magnetic field plays the role of the
chemical potential for the JW-fermions. The spin thermal
conductivity is given by
\begin{eqnarray}\label{eMFTKappa}
\kappa_s =  \frac{Na}{\pi}  \int_{0}^{\pi/a}{\frac{df_k}{dT} \varepsilon_k v_k \ell_{s,k}  dk},
\end{eqnarray}
where $N$ is the number of spins per unit volume, $v_k =
\hbar^{-1} d\varepsilon_k/dk$ is the velocity and $\ell_{s,k}$ the
mean free path of the spin excitations. The applicability of the
MFT JW-fermion model is demonstrated by the example of the
specific heat $C_s(T,B)$, which is given by the right-hand side of
Eq.~(\ref{eMFTKappa}) omitting $v_k \ell_{s,k}$ from the
integrand. In Fig.~\ref{CmKmMFT}~(a), the calculated values of
$C_s(B)$ are compared with the experimental data of
Ref.~\onlinecite{Ruegg08cm} for $T= 0.35$ and
1.53~K~\cite{values}. The good agreement between the calculated
and measured $C_s(B)$ is obvious. Similarly, magnetostriction and
thermal expansion of \hp\ are well described by the JW fermion
model \cite{Lorenz08}.

In order to compare the measured thermal conductivity with the
expected spin thermal conductivity, information about the mean
free path $\ell_s$ is required. At low temperatures, scattering by
magnetic impurities and disorder is dominant. The scattering by
phonons should be rather weak in the present case, because the
concentration of phonons is much lower than the concentration of
spin excitations, as the bandwidth of spinons ($\sim J_\|\simeq
3.6$~K) is much smaller than the phonon bandwidth (of the order of
the Debye temperature $\theta_{\rm D} \simeq 10^{2}$~K). In
previous studies of the heat transport in
spin-1/2-ladder~\cite{Hess06} and spin-1-chain
compounds~\cite{Sologubenko08} the impurity-limited mean free path
is found to be $(k,B,T)$-independent and of the order of the mean
distance between impurities in the ladder (chain) direction. Our
analysis of the low-temperature magnetization data of similarly
grown \hp\ crystals yields an average distance of about
0.16~$\mu$m between "defective" rungs, i.e.\ rungs that are not in
the singlet state for low fields. The corresponding calculated
minimum spin thermal conductivity $\kappa_{s, {\rm min}}(B)$ is
shown for $T=0.37$~K in Fig.~\ref{CmKmMFT}~(b) and compared with
the measured total $\kp$. Near the critical fields, where the gap
closes, $\kappa_s(B)$ should show a characteristic two-step
increase from a negligibly small value to a constant of the order
100~mW~m$^{-1}$~K$^{-1}$. Thus, one would expect an increase of
$\kappa_s$ that exceeds the measured total $\kappa = \kappa_s +
\kappa_{\rm ph}$ by more than one order of magnitude. Within the
experimental resolution (about 0.1~mW~m$^{-1}$~K$^{-1}$ at
$T=0.37$~K) our $\kappa(B)$ data do not show any indication of
such steps near either $B_{c1}$ or $B_{c2}$. The absence of the
expected steps cannot be simply explained by a very short mean
free path in the studied crystal, because $\kappa_s <
0.1$~mW~m$^{-1}$~K$^{-1}$ would correspond to an unphysically
small $\ell < 1$~\AA. Thus, the thermal conductivity in \hp\ is
dominated by the phononic contribution, $\kappa \approx
\kappa_{\rm ph}$. This conclusion is true for all temperatures we
studied.

The absence of $\kappa_s$ most likely results from impurity
scattering, but due to the extremely weak interladder coupling
this scattering is not consistent with a constant mean free path
given by the distance between impurities. As suggested in
Ref.~\cite{Sologubenko08}, disorder leads to a constant mean free
path of spin excitations only in the presence of weak yet
non-negligible interladder coupling. In particular, the condition
\begin{equation}\label{eDeloc}
        \bar{v}_s^\perp/a >  \bar{v}_s^\parallel/d_{\rm def}
\end{equation}
should be satisfied, where  $\bar{v}_s^\perp$ and $\bar{v}_s^\|$
are the characteristic velocities of, respectively, the spin
excitations perpendicular and parallel to the ladder direction,
and $d_{\rm def}$ is the average distance between the defects.
When Eq.~(\ref{eDeloc}) is satisfied, an energy transfer between
neighboring chains is possible and $\ell_s \approx d_{\rm def}$.
Otherwise, strong backscattering by impurities combined with the
low probability of interchain transfer may lead to a spinon
localization and thus $\kappa_s \approx 0$. In our recent heat
transport studies in the spin-chain compounds \CuPzN\ and NENP
\cite{Sologubenko07,Sologubenko08}, the left-hand side of the
inequality (\ref{eDeloc}) was an order of magnitude larger than
the right-hand side for the temperature range where a large $\l_s$
has been identified. However, in the present case of \hp\, for all
investigated temperatures, the estimates show that
$\bar{v}_s^\perp/a  \leq \bar{v}_s^\parallel/d_{\rm def}$.
Therefore, we conclude that the absence of a measurable spin
contribution to the heat transport results from a localization of
the spin excitations in finite ladder segments of \hp .

Theoretical efforts treating localization effects in 1D systems
have focused on the electrical conductivity
\cite{GiamarchiBook04}. Only recently random-disorder induced
localization in the spin thermal conductivity of the spin-1/2 XXZ
chain has been specifically addressed by numerical diagonalization
of small-size systems up to 20 spins \cite{Karahalios08cm}. For
finite $T$, these results suggest the possibility of spinon
localization for the easy-plane case ($\delta=0$ in
Eq.~(\ref{eMFTdisp})). Similar calculations, which would be
applicable for our case ($\delta =1/2$ and much longer $d_{\rm
def}$), have not been published so far.

\begin{figure}[t]
   \begin{center}
     \includegraphics[width= 8.6cm]{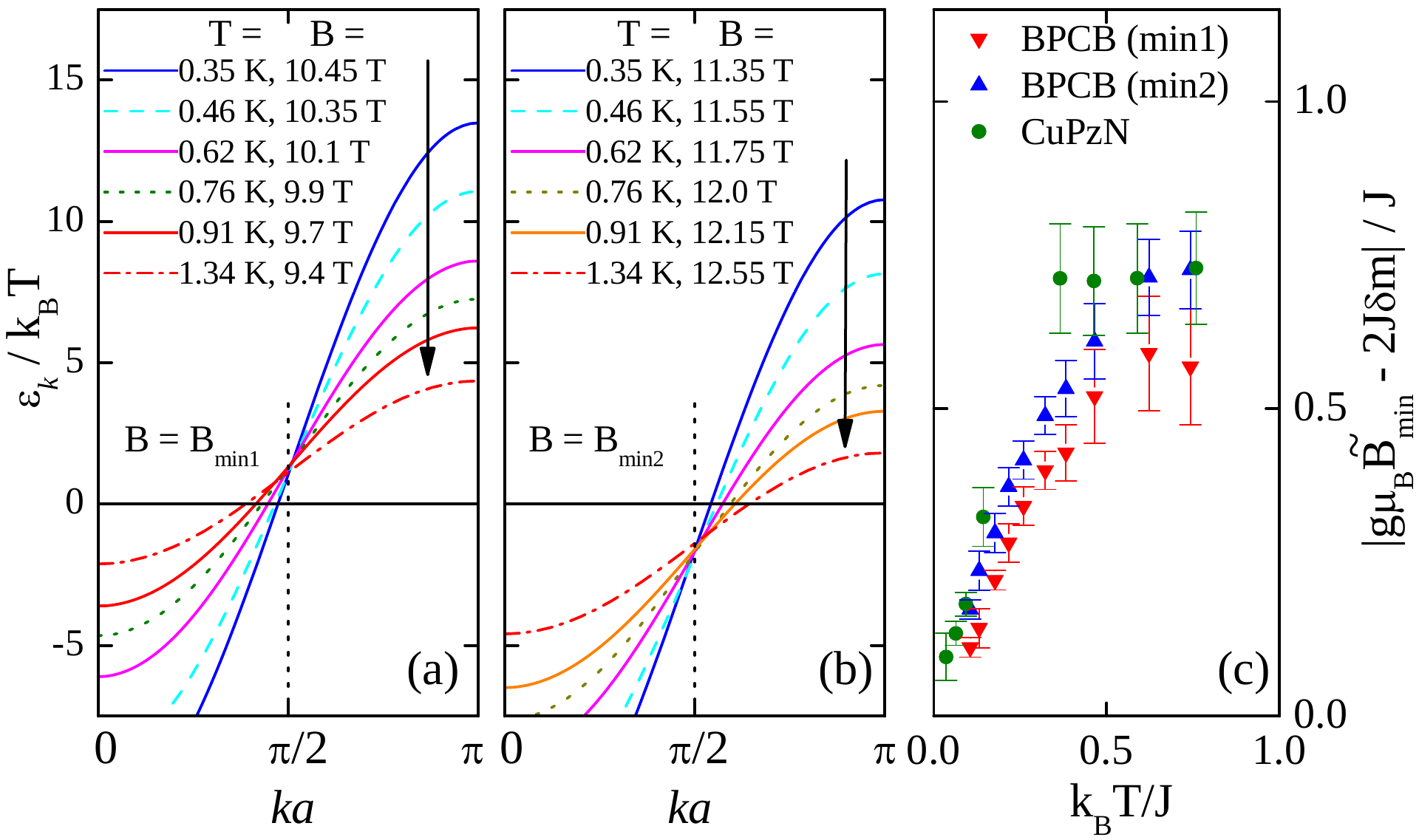}
      \caption{(color online)
     (a,b) Energy of the spin excitations normalized by $k_{\rm B}T$ as a function of the
     wavevector. The dispersion curves are calculated via Eq.~(\ref{eMFTdisp}) for $T$ and $B$
     corresponding to the minima of $\kappa(B)$. The arrows
     indicate increasing $T$ for $ka>\pi/2$.
    (c) Spin excitation energy at $ka=\pi/2$ for $B=B_{\rm min}$,
    calculated via Eq.~(\ref{eMFTdisp}), as a
    function of temperature for \hp\ (BPCB) and \CuPzN\ (CuPzN). Both
    scales are normalized by the respective exchange couplings $J$ (see text).  }
\label{KphB}
\end{center}
\end{figure}

Now we turn to another salient feature of the measured
$\kappa(B)$, namely, the two minima of $\kappa_{\rm ph}(B)$ in the
vicinity of $B_{\rm middle}$. These minima are obviously caused by
the scattering of phonons by spin excitations. In order to gain
insight into what part of the spinon spectrum contributes to the
scattering of phonons one may use the dominant-phonon method. At
low temperatures, when phonons are mainly scattered by sample
boundaries ($\ell_{\rm ph}={\rm const}$), phonons with energy
$\simeq 3.7\, k_{\rm B} T$ provide most of the total thermal
conductivity \cite{BermanBook}. Thus, any additional phonon
scattering mechanism which scatters phonons around a certain
frequency $\omega_r$, produces the strongest reduction of
$\kappa_{\rm ph}$ at $T \simeq \hbar \omega_r / 3.7 k_{\rm B}$.
Hence, if particular spinons (in $k$ space) are responsible for
the minima of $\kappa(B_{\rm min},T)$, the energy of these spinons
$\varepsilon_k(B_{\rm min},T)$ should scale linearly with
temperature. Accordingly, we calculated the dispersion curves
$\varepsilon_k$ of the JW fermions for the $(B,T)$ values of the
lower ($B=B_{\rm min1}$) and upper ($B=B_{\rm min2}$) minimum via
Eq.~(\ref{eMFTdisp}) and normalized them by $k_{\rm B}T$. The
result is shown Fig.~\ref{KphB}~(a,b). All calculated curves
$\varepsilon_{k}/k_{\rm B}T$ practically meet at the commensurate
point $ka=\pi/2$, suggesting that the spinons with $ka \simeq
\pi/2$ are responsible for this additional phonon scattering.

This type of scattering seems not to be restricted to \hp. Minima
of $\kappa(B)$ have also been observed in the spin-1/2 chain
material \CuPzN\ \cite{Sologubenko07}. Fig.~\ref{KphB}(c) compares
the energies of the $ka=\pi/2$ JW-fermions
$\varepsilon_{\pi/2a}(B_{\rm min}) = | - g \mu_B \tilde{B}_{\rm
min} + 2 J \delta m |$, Eq.~(\ref{eMFTdisp}), as a function of
temperature for \hp\ and for \CuPzN. Note, that both axes have
been normalized to $J$ of the respective compound and that
$\tilde{B} \equiv B$ for the spin-1/2 chains. For both materials
we observe essentially the same behavior in the LL state
suggesting a common and universal origin of this additional phonon
scattering, which is most likely associated with Umklapp
scattering. As discussed in Ref.~\onlinecite{Rasch09}, spinon
Umklapp scattering in the presence of weak disorder can lead to
minima in the field dependence of the {\it spin} thermal
conductivity $\kappa_s(B)$. However, the minima discussed in the
present paper are observed in the {\it phonon} part of the heat
conductivity $\kappa_{\rm ph}$.

In conclusion, our measurements of the thermal conductivity in the
two-leg spin-1/2 ladder compound \hp\ provides clear evidence for
the absence of spin-mediated heat transport. The most likely
origin is a spinon localization in finite ladder segments, which
is favored by the extremely weak interladder coupling, i.e.\ the
high degree of one-dimensionality, in \hp. The strongest effect
observed in our experiments is a suppression of the phonon heat
transport due to phonon-spinon scattering, which is exceptionally
strong close to the commensurate filling of the spin excitation
band.

 \acknowledgments
We thank A. Rosch for many stimulating discussions and a critical
reading of the manuscript. The contribution of H.~R. Ott at the
initial stage of the project is appreciated. We acknowledge useful
discussions with C. Batista, M. Garst, P. Prelov\v{s}ek,  E.
Shimshoni. This work was supported by the Royal Society, by the
Swiss National Science Foundation, and by the Deutsche
Forschungsgemeinschaft through SFB~608.


\begin{thebibliography}{10}

\bibitem{HeidrichMeisner07}
F. Heidrich-Meisner, A. Honecker, and W. Brenig, Eur. Phys. J. Special Topics
  {\bf 151},  135   (2007).

\bibitem{Sologubenko07_JLTP}
A.~V. Sologubenko {\em et al.},
J. Low Temp. Phys.
  {\bf 147},  387  (2007).

\bibitem{Hess07_EPJ}
C. Hess, Eur. Phys. J. Special Topics {\bf 151}, 73 (2007).

\bibitem{Sologubenko00_lad}
A.~V. Sologubenko {\em et al.},
  Phys. Rev. Lett. {\bf 84},  2714  (2000).

\bibitem{Hess01}
C. Hess {\em et al.},
  Phys. Rev. B {\bf 64},  184305  (2001).

\bibitem{Kudo01}
K. Kudo {\em et al.},
 J. Phys. Soc. Jpn. {\bf 70},  437  (2001).

\bibitem{Patyal90}
B.~R. Patyal, B.~L. Scott, and R.~D. Willett, Phys. Rev. B {\bf 41},  1657
  (1990).

\bibitem{Watson01}
B.~C. Watson {\em et al.},
 Phys. Rev. Lett. {\bf 86},  5168  (2001).

\bibitem{Lorenz08}
T. Lorenz {\em et al.},
 Phys. Rev. Lett. {\bf 100},  067208  (2008).

\bibitem{Klanjsek08}
M. Klanj\v{s}ek {\em et al.},
  Phys. Rev. Lett. {\bf 101},  137207  (2008).

\bibitem{Anfuso08}
F. Anfuso {\em et al.},
 Phys. Rev. B {\bf 77},  235113  (2008).

\bibitem{Ruegg08cm}
Ch. R\"uegg {\em et al.},
 Phys. Rev. Lett. {\bf 101}, 247202 (2008) .

\bibitem{Thielemann08cm}
B. Thielemann {\em et al.},
 Phys. Rev. B {\bf 79} , 020408(R) (2009)

\bibitem{Thielemann09PRL}
B. Thielemann {\em et al.},
 Phys. Rev. Lett. {\bf 102},  107204  (2009).

\bibitem{Boulat07}
E. Boulat {\em et al.},
 Phys Rev. B {\bf 76},  214411  (2007).

\bibitem{Totsuka98}
K. Totsuka, Phys. Rev. B {\bf 57},  3454  (1998).

\bibitem{Mila98}
F. Mila, Eur. Phys. J. B {\bf 6},  201  (1998).

\bibitem{Chaboussant98}
G. Chaboussant {\em et al.},
 Eur. Phys. J. B {\bf 6},  167 (1998).

\bibitem{HeidrichMeisner05}
F. Heidrich-Meisner, A. Honecker, and W. Brenig, Phys. Rev. B {\bf 71},  184415
   (2005).

\bibitem{HeidrichMeisner05PhD}
F. Heidrich-Meisner, PhD thesis, Technische Universit\"{a}t Braunschweig, 2005.

\bibitem{values}
The calculations were done with the parameters $g=2.06$,
$J_\parallel$ = 3.5~K, $J_\perp$ = 13.2~K, $N=1.120\times
10^{27}$~m$^{-3}$ (number of rungs per unit volume), and
$a=8.5$~\AA .

\bibitem{Hess06}
C. Hess {\em et al.},
 Phys. Rev. B {\bf 73},  104407  (2006).

\bibitem{Sologubenko08}
A.~V. Sologubenko {\em et al.},
 Phys. Rev. Lett. {\bf 100},  137202  (2008).

\bibitem{Sologubenko07}
A.~V. Sologubenko {\em et al.},
 Phys. Rev. Lett. {\bf 98},  107201  (2007).

\bibitem{GiamarchiBook04}
T. Giamarchi, {\em Quantum Physics in One Dimension} (Oxford University Press,
  Oxford, 2004).

\bibitem{Karahalios08cm}
A. Karahalios {\em et al.},
 Phys. Rev. B  {\bf 79}, 024425 (2009).

\bibitem{BermanBook}
R. Berman, {\em Thermal conduction in solids} (Clarendon Press, Oxford, 1976).

\bibitem{Rasch09}
E. Shimshoni {\em et al.},
 Phys. Rev. B  {\bf 79}, 064406 (2009).

\end{thebibliography}
\end{document}